\documentclass[12pt]{iopart}
\usepackage{graphicx}
\begin{document}

\title[Bag vs. NJL models for colour-flavour-locked strange quark matter]{Bag vs. NJL models for colour-flavour-locked strange quark matter}

\author{L Paulucci$^1$, E J Ferrer$^2$, J E Horvath$^3$ and V de la Incera$^2$}
\address{$^1$Universidade Federal do ABC, Av. dos Estados, 5001, 09210-580, Santo Andr\'e, SP, Brazil}
\address{$^2$Department of Physics, University of Texas at El Paso, El Paso, TX 79968, USA}
\address{$^3$Instituto de Astronomia, Geof\'\i sica e Ci\^encias Atmosf\'ericas, Rua do
Mat\~ao 1226, 05508-900, S\~ao Paulo, SP, Brazil}

\eads{\mailto{laura.paulucci@ufabc.edu.br},\mailto{ejferrer@utep.edu},\mailto{foton@astro.iag.usp.br},\mailto{vincera@utep.edu}}

\begin{abstract}
 We compare the mass-radius relationship of strange stars obtained in two theoretical frameworks describing the
colour-flavuor-locking state of dense quark matter: The semi-empirical MIT model and a self-consistent approach using
the Nambu-Jona-Lasinio (NJL) model. In the simplest MIT model extended to include pairing, one can make the equation
of state stiffer by increasing the gap parameter so that larger maximum masses for these objects can be reached. In
the NJL model, however, such an effect is not possible. To increase the gap parameter within the NJL model to values
comparable to those considered in the MIT case, a noticeably increase of the diquark-coupling-constant strength
is needed, but this in turn {\it softens} the equation of state producing a lower maximum star mass. This behaviour
is interpreted as signalling the system crossover at high diquark coupling from a BCS regime to a BEC one, a process
that cannot be reproduced within the simple MIT prescription.
\end{abstract}

\pacs{21.65.Qr, 26.60.Kp, 97.60.Jd}

\submitto{\JPG}

\maketitle

\section{Introduction}

The exact composition of neutron stars is still under debate, with proposals ranging from nuclear matter (possibly with
hyperons and superfluid nucleons) to deconfined quark matter (either two or three flavours). New data on masses and
radii, as well as the modelling of other phenomena (glitches, cooling, bursts episodes, etc), help constraining the
equation of state of matter in their interior, but no firm conclusion has been drawn yet.
There are a few parameters to be adjusted in both nuclear and quark descriptions, which should be further constrained
by nuclear matter data, but in any case their behaviour at ``zero temperature'' and large chemical potential remains uncertain.

The proposal that matter composed of up, down, and strange quarks, the so called strange quark matter (SQM), may have
a lower energy per baryon number
than the nucleon, thus being absolutely stable, dates back to the late 1970's \cite{Bodmer,Chin,Terazawa,Wit}.
Further developments raised the idea that color superconductivity should be the favoured state of SQM
since the superconducting gap would lower the total energy per baryon number of the system
\cite{Pairing1,Pairing2,Pairing3, German}. Since then much work has been done in order to characterise
such systems and to determine the parameters suitable for absolute and meta-stability.

Two frameworks have been mainly used in the study of SQM, either with or
without pairing between quarks: the
Nambu-Jona-Lasinio (NJL) and MIT bag models. They both present features which are in agreement with our current
understanding of nuclear matter, despite being inadequate to incorporate some known aspects of strongly interacting systems.

The MIT bag model was proposed in \cite{MIT} as a phenomenological model for explaining hadrons. Within this approach,
QCD is asymptotically free and confinement is achieved through the introduction of a vacuum pressure, the bag constant $B$,
that artificially maintains quarks inside a finite region in space. This model can also be applied to deconfined quark
matter in bulk, rendering the following thermodynamical potential for non-superconducting SQM at zero temperature and
strong constant coupling (it has been shown in \cite{Farhi84} that a finite $\alpha_c$ can be absorbed as an effective
reduction in $B$) for massless quarks

\begin{equation}\label{Omega}
\Omega=\sum_i\Omega_i+B,
\end{equation}
where
\begin{equation}\label{Omega1}
\Omega_i=-\frac{\mu_i^4}{4\pi^2},
\end{equation}
with $i$ running for  quarks u, d, s and electrons, and $\mu_i$ is the
chemical potential of particle $i$. Together with charge neutrality and chemical equilibrium conditions, the full features
of this phase are determined.

When considering pairing in the most symmetrical state, the colour-flavour-locked (CFL) phase of colour superconductivity
is realised at sufficiently high density \cite{Alford04}. In this situation, it is usual to use a semi-empirical model
in which the thermodynamical potential is assumed to be the sum of the one corresponding to the unpaired state (\ref{Omega}),
plus the gap ($\Delta$) depending leading term \cite{Rajagopal2001, Alford2001}, reading
\begin{equation}
\Omega_{CFL}^{MIT}=\sum_i\Omega_i-\frac{3}{\pi^2}\Delta_{CFL}^2\mu^2+B\label{Omega_MIT}
\end{equation}
The term $\sum_i\Omega_i$ represents a fictitious non-paired
state in which all quarks have a common Fermi momentum and the extra
term dependent on $\Delta$ represents the binding energy of the diquark
condensate.

On the other hand, in the NJL model, when neglecting quark masses in a CFL phase, the thermodynamic potential of this
phase is calculated to be \cite{Paulucci}
\begin{eqnarray}
\Omega_{CFL}^{NJL} =-\frac{1}{4\pi^2}\int_0^\infty dp p^2 e^{-p^2/\Lambda^2}(16|
\epsilon|+16|\overline{\epsilon}|)+ \nonumber \\
-\frac{1}{4\pi^2}\int_0^\infty
dp p^2 e^{-p^2/\Lambda^2}(2|\epsilon'|+2|\overline{\epsilon'}|)+
\frac{3\Delta_{CFL}^2}{G}+B \label{OmegaCFL}
\end{eqnarray}
where
\begin{eqnarray*}\label{6}
\varepsilon=\pm \sqrt{(p-\mu)^2+\Delta_{CFL}^2}, \quad
\overline{\varepsilon}=\pm \sqrt{(p+\mu)^2+\Delta_{CFL}^2}\nonumber
\\
\varepsilon'=\pm \sqrt{(p-\mu)^2+4\Delta_{CFL}^2,}\quad
 \overline{\varepsilon'}=\pm \sqrt{(p+\mu)^2+4\Delta_{CFL}^2}
\end{eqnarray*}
are the quasiparticles dispersion relations. In order to have
only continuous thermodynamical quantities, we
introduced in (\ref{OmegaCFL}) a smooth cutoff depending on the effective-theory energy scale $\Lambda$.

In astrophysical applications, when finding the stellar mass-radius relation, one needs to obtain the equation of state (EoS),
a relation between energy density and pressure of matter. These quantities can be obtained for an isotropic system 
from the thermodynamic potential through the relations
\begin{equation}  \label{Energy-Density}
\epsilon_{CFL}=\Omega_{CFL}-\mu \frac{\partial \Omega_{CFL}}{\partial \mu},
\end{equation}
\begin{equation}  \label{Prssure}
P_{CFL}=-\Omega_{CFL}
\end{equation}
where $\Omega_{CFL}$ is evaluated in the solution of the gap equation $\frac{\partial \Omega_{CFL}}{\partial\Delta_{CFL}}=0$
in the NJL approach, whereas $\Delta$ is given and fixed by hand in the MIT bag model.

\section {Results and Discussion}

 The particular observation for the pulsar J1614-2230, a binary system for which the mass of the neutron star was measured rather
accurately through the Shapiro delay, yielding M = 1.97$\pm$0.04 $M_{\odot}$ \cite{Demorest}, posed an important question about
the existence of strange stars. Of course, the answer to this question is directly related to the EoS derived from the model
used to describe strange matter. Moreover, the exact nature of the EoS will also influence other phenomena of the star like
cooling, glitches, burst episodes, etc. In this regard, we will compare and discuss the results obtained from the two previously
introduced approaches: the MIT bag model and the NJL model. These simple models
are the most widely used when considering the possible presence of SQM inside
compact objects, therefore it is important to gain a proper vision of their
behaviour, similarities and differences in treating quark matter.

\begin{figure}
\begin{center}
\includegraphics[width=0.5\textwidth]{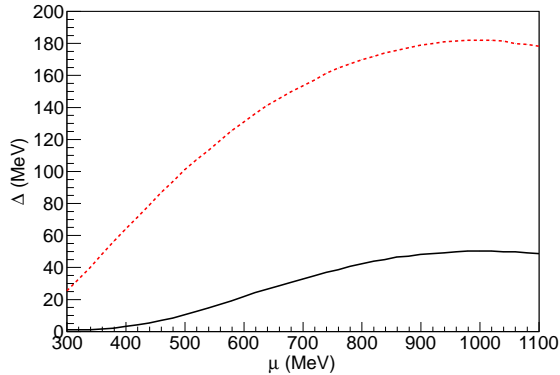}
\caption{\footnotesize Gap parameter behaviour with the chemical potential for CFL matter in the NJL theory with zero quark masses
and G=4.32 GeV$^{-2}$ (full line) and G=7.10 GeV$^{-2}$ (dashed line).} \label{Gap}
\end{center}
\end{figure}

An important part of the comparison of the models (\ref{Omega_MIT}) and (\ref{OmegaCFL}) is related to the assumptions 
about the pairing terms.
In the MIT bag model the value for the gap parameter is fixed by hand, hence it does
not explicitly depend on changes in other parameters characterising the mixture.
However, as pointed above, in the NJL approach the pairing gap is obtained through the gap equation
$\partial \Omega / \partial \Delta = 0$. In this way, $\Delta$ is dependent on the density and diquark coupling constant $G$,
as can be seen in figure \ref{Gap}.

\begin{figure}
\begin{center}
\includegraphics[width=0.49\textwidth]{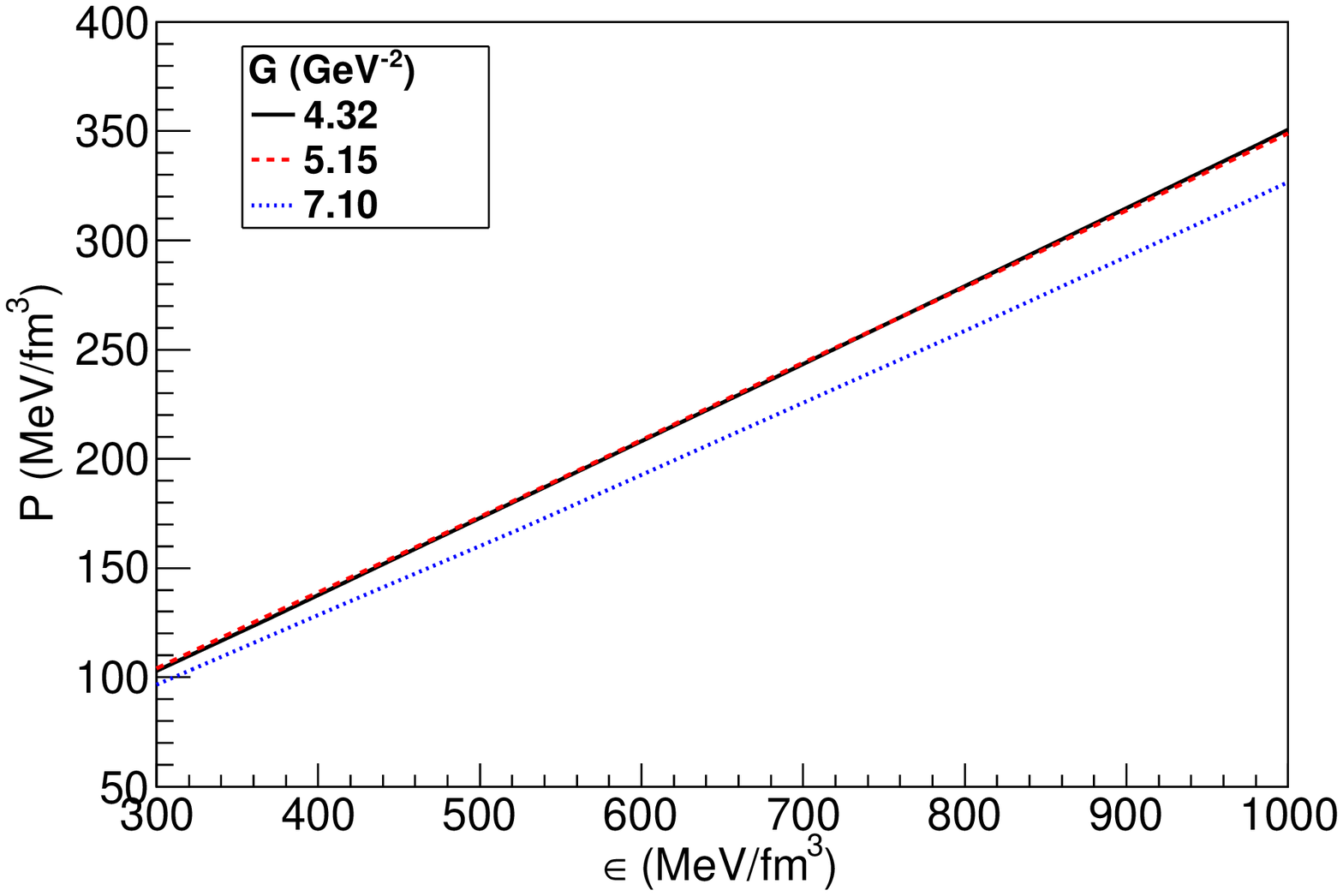}
\includegraphics[width=0.49\textwidth]{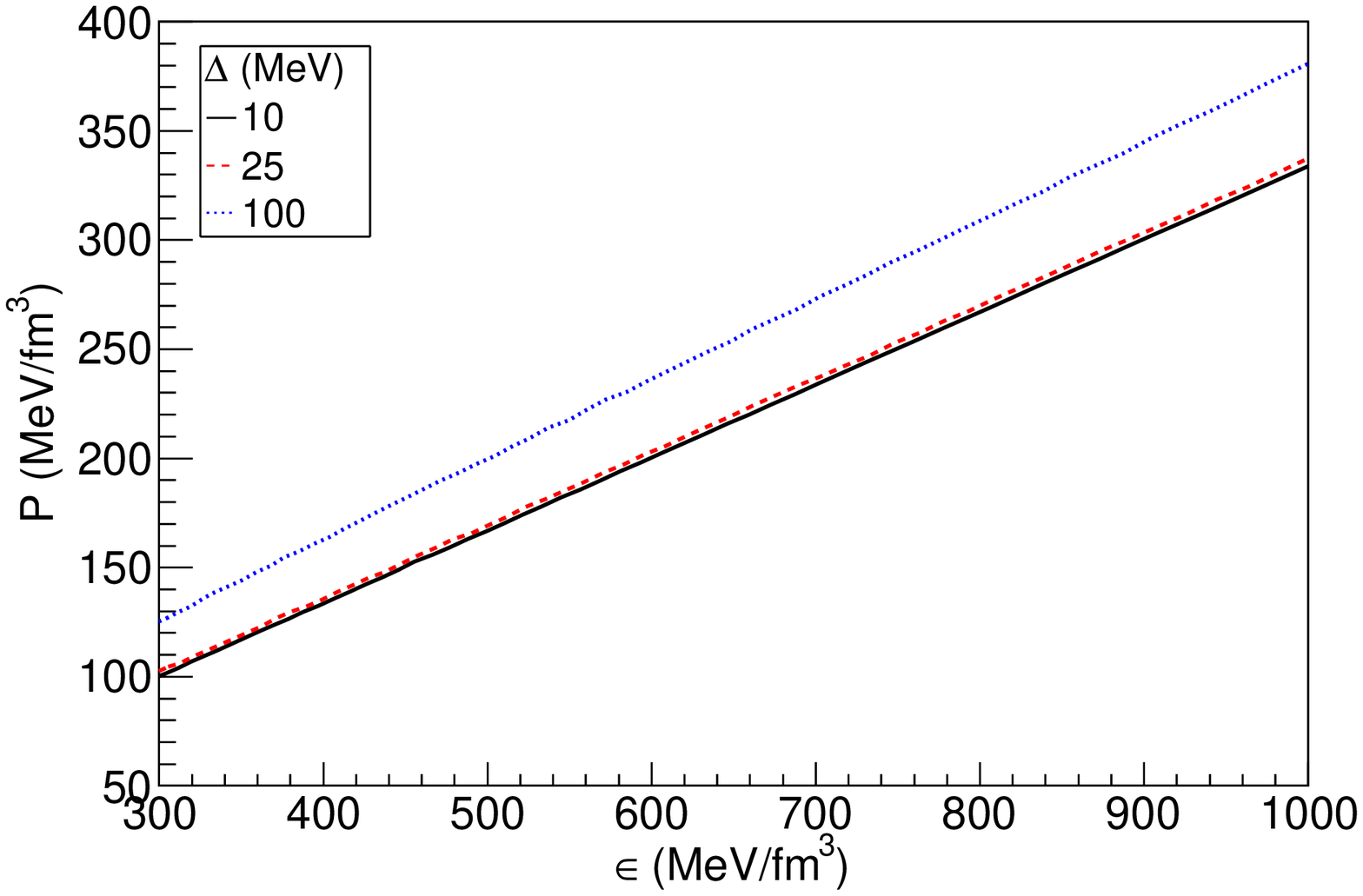}
\caption{\footnotesize In the left panel, equation of state for CFL matter within the NJL theory for different values of G. In the right
panel, the same for the MIT bag model for different values of $\Delta$. They consider zero quark masses and do not show the influence of
the bag constant.} \label{EoS}
\end{center}
\end{figure}

Using coupling constant values  G = 4.32, 5.15, and 7.10 GeV$^{-2}$, corresponding to $\Delta$ = 10, 25, and 100 MeV at $\mu=500$ MeV 
respectively,
the equation of state in the NJL theory in the region of interest for strange stars is shown in figure \ref{EoS}, as well as the EoS for
CFL matter in the MIT bag model for two different values of the gap parameter. It is important to notice however, that G = 7.10 GeV$^{-2}$
was used with the only purpose of comparing the NJL results with the MIT ones, because a gap $\Delta= 100$ MeV is of common use in the MIT model.
It can be observed that the splitting between the EoS for different $\Delta$'s is more significant in the MIT model than in the NJL one.

It can be easily checked for the MIT results that the higher the gap, the stiffer the EoS. Hence, as the mass supported by a given star
configuration is related to the stiffness of the EoS, a higher value of the gap renders higher maximum masses for stable strange stars \cite{German}.

Note, however, that this is {\it not} the case for the NJL calculations. When a higher
value of $G$ is used, although the corresponding gap parameter increases for each value of $\mu$, the EoS does not change considerably
and actually weakly {\it softens} in the region of interest for compact star interiors. Therefore, in the NJL approach it is not possible to
increase the maximum mass that can be supported by strange stars, even when unphysical large values of the coupling constant are employed.

The origin of the softening of the EoS in the CFL-NJL is due to the term $3\Delta^2/G$ that enters with a negative sign in the pressure.
Notice that the softening of the EoS with stronger interaction does not occur in the 2SC case \cite{PRD69}. This apparent contradiction
can be understood after realising that in the CFL case there is a factor of 3 in the $\Delta^2/G$, but that factor becomes 1/4 in the 2SC
case \cite{NPA}, so this term does not affect the pressure as much as in the CFL case.

From a physical point of view, the increase of $G$ beyond a certain value in the CFL-NJL model implies a softening of the EoS because for
large enough $G$'s the system begins to crossover from  BCS to BEC \cite{BCS-BEC1, BCS-BEC2, BCS-BEC3}. The crossover is reflected in the
decrease of the system pressure, which is due to an increment in the number of diquarks that become Bose-like molecules and hence cannot
contribute to the dominating Pauli pressure of the system. As shown in \cite{Jason}, if the diquark coupling is high enough to produce
the crossover from the BCS to the BEC regime, the pressure of the system formed now by Bose-like molecules at zero temperature
would become zero, signalling an instability in the stellar system. For the CFL phase, an identical behaviour is found. The value of the critical coupling for the crossover in this case is $G_{cr}\approx 7.2$GeV$^{-2}$ \cite{Israel}.

On the other hand, in the semi-empirical MIT model approach,
the derivation of (\ref{Omega_MIT}) \cite{Schafer, Evans}
is made under the assumption that an expansion in the small parameter $\Delta/\mu$ is valid, requiring either densities much higher than
those expected in the compact
star interiors, or constraining the gap magnitude to relatively small values. Working in the
regime of weak coupling, and assuming a small four-fermion coupling, gives the correct order
of magnitude of the pairing gap, but one must keep in mind that the gap full dependence
on the density cannot be disregarded in favour of an arbitrary constant value.
Within the MIT model, if one has to put by hand large values of $\Delta$ to reach high stellar masses,
it means working in regions of
the parameter space in which $\Delta/\mu$ is not necessarily small any more. Therefore, it seems that the NJL model, where the gap is always
found self-consistently as a function of the chemical potential is a much more reliable approach to explore the mass-radius curves. In
both cases, however, the important question of the confinement remains unsolved.

Recent data \cite{Ozels1, Ozels2, Demorest}
have determined masses and/or radii for some compact objects with high precision (although some of these remain to be confirmed \cite{Lattimer}),
rendering some information about the composition of these objects (see, for example, \cite{AHP}).
In addition, new data on black widow systems
\cite{Heavy1, Heavy2} suggests the existence of even more massive compact stars ($M = 2.1 - 2.9 M_{\odot}$).

\begin{figure}
\begin{center}
\includegraphics[width=0.49\textwidth]{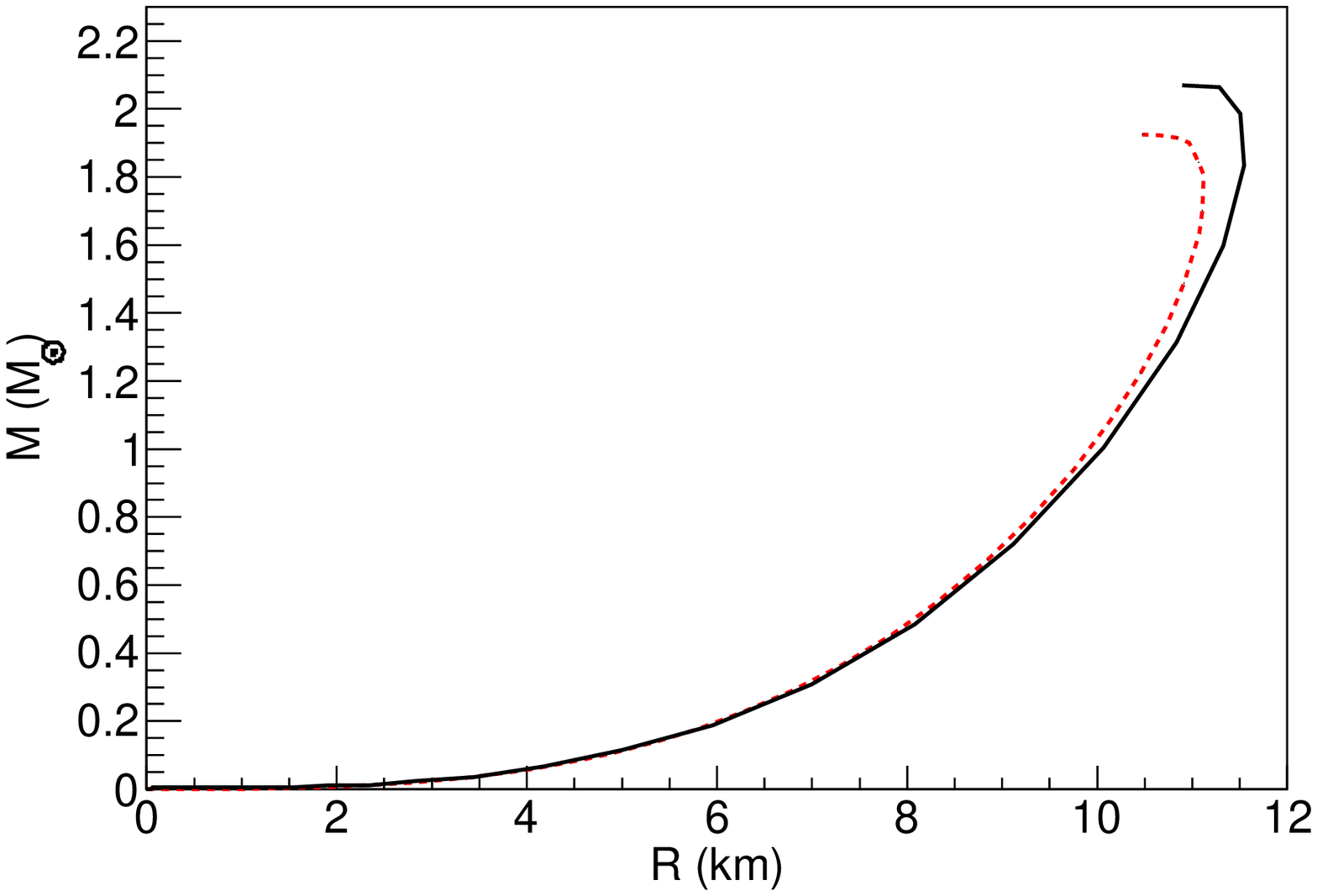}
\includegraphics[width=0.49\textwidth]{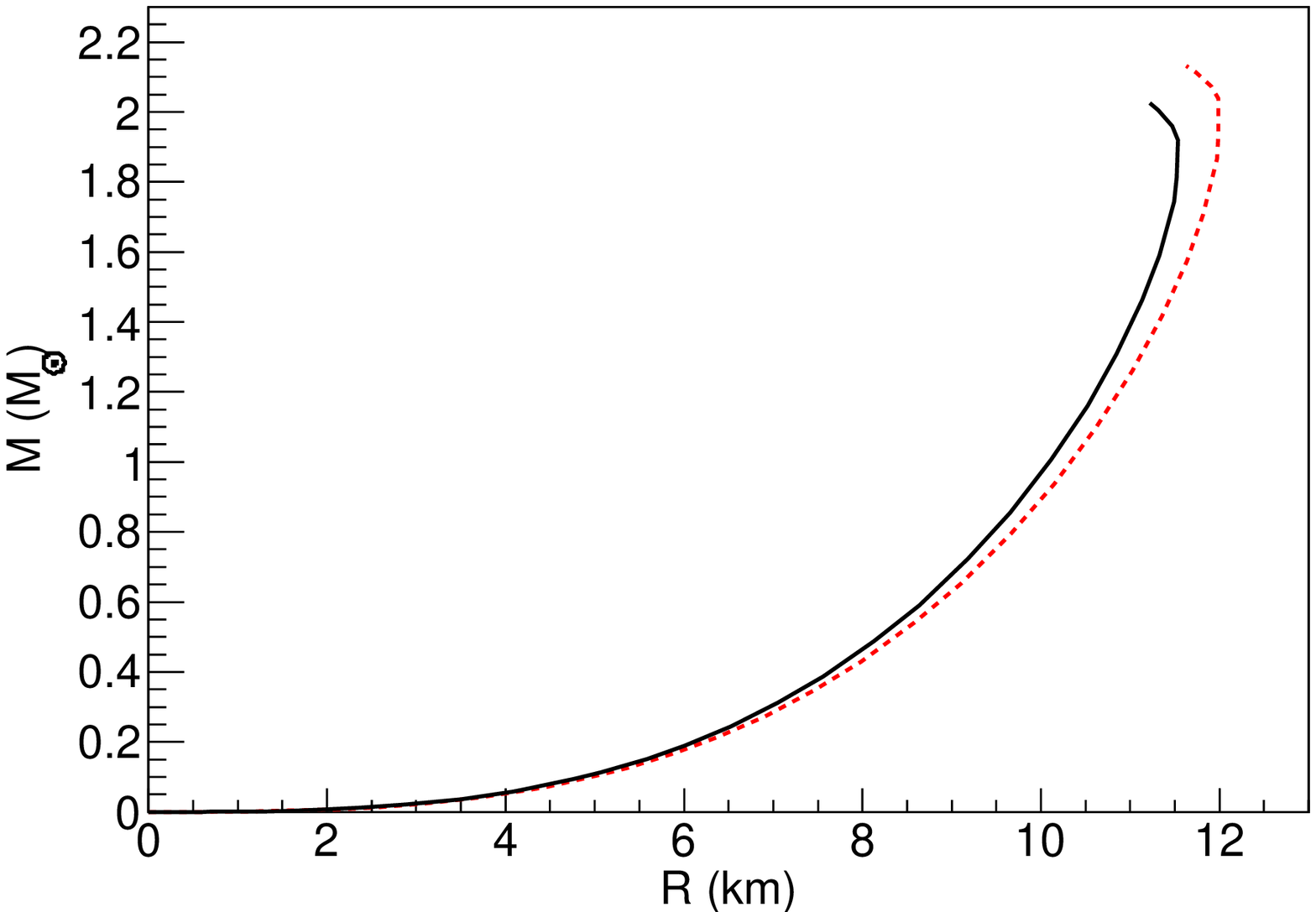}
\caption{\footnotesize Mass-radius relation for CFL matter with $B$=58 MeV/fm$^3$. Results obtained with NJL theory in the left panel:
full line for G=4.32 GeV$^{-2}$ and dashed line for G=7.10 GeV$^{-2}$. Results obtained with MIT bag model in the right panel with
$\Delta$ = 10 (full line) and 100 MeV (dashed line).} \label{MR}
\end{center}
\end{figure}

If these values are confirmed with small error bars, neither one of those current models would be able to
explain observations. This conclusion can be taken from the analysis of bare strange stars with ``optimal'' parameters, i.e., by taking parameters that describe the stiffest EoS for strange quark matter within a given model, as in figure \ref{MR}. Even if the star has a normal nuclear surface, as inferred from X-ray burst episodes, the size of the crust should be small enough not to influence significantly the mass-radius relation obtained for a bare strange star. In the original work of Alcock et al. \cite{Alcock86} the maximum crust that could be supported by a strange star with mass $\sim 1.4 M_{\odot}$ is estimated to have a mass of $\sim 10^{-5} M_{\odot}$ and thickness of $\sim 200$ m. This should be enough to support bursts 
in accreting X-binaries.
The maximum mass in the NJL approach is bounded from above, since any increase in the values of the parameters $B, m_s, G$
will produce a smaller maximum mass, as reflected in figure \ref{MR} for changes in the value of the coupling $G$. The MIT model could in principle account for these high masses by increasing the gap value, but as
pointed out above, this stiffening of the EoS is artificial and does not represent the physical behaviour of the system, which tends to
crossover form a BCS to a BEC regime, and the validity of the expansion used in the MIT model becomes questionable.


\section{Concluding Remarks}

We have shown that the two widely used models for describing superconducting
quark matter in the CFL state do not behave in a similar way with respect to changes in
the pairing gap of the system. The equation of state does not stiffen in the
NJL model because the increase of the gap in this case comes from
strengthening the diquark coupling constant $G$, which in turn favours the
crossover to the BEC regime and tends to decrease the pressure \cite{Jason}.
This is at odds with the approach within the MIT bag model used for example
in \cite{Alford2001, German}, in which a higher $\Delta$ stiffens the EoS, rendering a higher 
maximum mass for strange stars.

This indicates that in the simple model (\ref{Omega_MIT}), the contribution of the pairing energy to an effective vacuum
that pushes the maximum masses to high values should be bounded from above, as can be corroborated by using a self-consistent
approach like the NJL-CFL model. Very high values of $\Delta$ should not be employed in stellar calculations.

The approximation made in Refs. \cite{Schafer, Evans}
cannot be blindly applied to the density range important for neutron star physics; therefore, the conclusions drawn
about the compatibility of recent data for maximum neutron star masses with a given EoS (e. g. \cite{AHP}),
as well as any automatic assumption that a higher value for the gap parameter renders a higher maximum mass and other calculations made employing very high values of the gap parameter within the MIT bag model
(e. g. \cite{Delta1, Delta2, Delta3}) should be revisited.

On a final note, it is worth to mention that an extended NJL model that includes vector interactions is a better candidate to stiffen the EoS (\cite{Vector-vector}). Nevertheless, this effect is dependent on the unknown value of the new interaction coupling.

\ack{The authors acknowledge the financial support received from
Funda\c c\~ao de Amparo \`a Pesquisa do Estado de S\~ao Paulo and
from the CNPq Agency (Brazil). The work of EJF and VI was supported in part by the Office of Nuclear Theory of
the Department of Energy under contract de-sc0002179.}

\bibliographystyle{unsrt}

\bibliography{njl_mit}

\end{document}